\documentclass[12pt,fleqn]{article}
\usepackage{amsmath,amssymb}
\begin {document}
\title{ {\Large Linear algebra of reduced units and discussion of temperature parameter}}
\author{Christopher G. Jesudason\footnote{Correspondence: Christopher G. Jesudason,Chemistry Department, University of Malaya, 50603 Kuala Lumpur, Malaysia; E-mail: jesu@um.edu.my}\\
{\normalsize Chemistry Department, University of Malaya,}\\ [1 mm]
{\normalsize 50603 Kuala Lumpur, Malaysia}\\
}
\date{{\normalsize 23 February, 2004} }
%\address{Chemistry Department
%University of Malaya,50603 Kuala Lumpur,Malaysia.}
\parindent=0mm
\maketitle
\newtheorem{defn}{Definition}

\newtheorem{rem}{Remark}
\newtheorem{axm}{Axiom}
\newcommand{\db}[1]{\mathbf{d_{#1}} }
\newcommand{\Mb}[1]{ {\left[\,\prod_{i=1}^{m}\mathrm{M_i}^{#1}\,\right]}}
\newcommand{\M}[1]{ {\left[\mathrm{M}{#1}\,\right]}}
\newcommand{\Mbl}[2]{ {\left[\,\prod_{i=1}^{m}\mathrm{M_{#2}}^{#1}\,\right]}}
\newcommand {\Fi}[3] {F_i^{\alpha,\,\beta,\ldots \, \gamma}\left(#1,\,#2\,\ldots #3\,\right)}
\newcommand {\fisca}[4] {f_i^{#1}(#2,\,#3,\,\ldots #4)}
\newcommand {\MLT}[3] {\left[\mathrm{M^{#1} L^{#2} T^{#3}} \right]}
\newcommand {\Piprod}[1] {\prod_{i=1}^{m{#1}}}

\newcommand {\Fidwn}[4] {F_{#4}^{\alpha,\,\beta,\ldots \, \gamma}\left(#1,\,#2\,\ldots #3\,\right)}

\newcommand {\T}{\mathcal{T}}
\newcommand{\R}{{\mathbb R}}
 
\newcommand{\Z}{{\mathbb Z}}

\abstract{\noindent A formal linear vector field  representation for scientific equations is developed to rationalize the intuitive methods that are constantly employed. It is shown that unlike mechanical units that appear in  the basis of the space, the reduced temperature and  Boltzmann parameter cannot be described by the basis set individually and can only be described as a product. Further, the definition  and determination of  temperature is dependent on theory and not on standard mechanical units. It is shown that there is no reason to reduce the number of degrees of freedom in temperature determination via equipartition since stochastic variables are involved, and this observation  is significant in that the temperature  variable  reported  in simulation studies would have a discrepancy  to the extent of using the  decreased number of freedom, which is most cases is not large nor significant.   The standard assignments used in reduced units  do not lead to errors because operationally the resulting reduced temperature parameter  represents the reduced product of the temperature  and Boltzmann parameters. The non-independence of these quantities explains   why entropy and other associated functions cannot be calculated directly, but are always scaled in dimensionless increments of the Boltzmann parameter}
 
\section {Introduction to vector field properties of scientific equations }
\noindent 
A scientific  result  of magnitude $c_i$ may be represented as  $c_{i}.\left[\prod_{i=1}^{m}\mathrm{M_i^{\alpha_i}}\right]$ 
where usually  $\alpha_i\in \Z ,\, c_i \in F \,$ for field $F$  and where normally the restriction $c_i \in \R$ is utilized
only but there is no reason why $c_i$  or $\alpha_i$ cannot belong to a larger (e.g. complex) field. The $\mathrm{M_i}$ are the so-called fundamental units; e.g.  $c.\left[\mathrm{M^\alpha L^\beta T^\gamma} \right]$ may describe  a quantity of magnitude $c$ characterized by exponents $(\alpha,\beta,\gamma)$ in mass (M), length (L) and time (T) respectively. The set $\left\{\mathrm {M_i}\right\}$, in general, and in particular  ${\mathrm {M,\,L,\,T}}$ is isomorphous to any fixed members  in $F$ where the exponents of products (operator $\otimes$) is concerned, i.e. $\left[\,\prod_{i=1}^{m}\mathrm{M_i}^{\alpha_i}\,\right]\otimes \left[\,\prod_{i=1}^{m}\mathrm{M_i}^{\beta_i}\right]=\left[\prod_{i=1}^{m}\mathrm{M_i}^{(\alpha_i + \beta_i)}\right]$. Let $\mathbf{d_{\alpha}}=\left[\,\prod_{i=1}^{m}\mathrm{M_i}^{\alpha_i}\,\right]$ define the "`unit dimension"' or basis  of the scientific quantity; then $\mathbf{d_{\alpha}}$ is uniquely characterized by the m-tuple $(\alpha_1,\, \alpha_2,\, \ldots \, \alpha_m)$. Define the zero element  and unit element of $\mathbf{d_{\alpha}}$ as $0.\left[\,\prod_{i=1}^{m}\mathrm{M_i}^{\alpha_i}\,\right]$ and $1.\left[\,\prod_{i=1}^{m}\mathrm{M_i}^{\alpha_i}\,\right]$ respectively. We define  the addition operator $\oplus$ such that 
\begin{equation} \label{e1}
c_{\gamma_1}.\mathbf{d_{\alpha}} \, \oplus  \,   c_{\gamma_2}.\mathbf{d_{\alpha}}\mathbf
=(c_{\gamma_1}+c_{\gamma_2}).\mathbf{d_{\alpha}}
\end{equation}
The scalar multiplication of element $c_{\alpha_1}.\db{\alpha}$  by $q\in F$ is defined as 
\begin{equation} \label{e2}
	q(c_{\alpha}.\db{\alpha})=(c_{\alpha_1}.q).\db{\alpha}.
\end{equation}

 For any vectors $\mathbf {\alpha,\,\beta\,,\gamma}$ where $\mathbf{\alpha_{i}}=\alpha_{i}.\db{\alpha}$ in space $\db{\alpha}$, the operator $\oplus$ is commutative, associative, and properties (1) and (2) ensures that $\mathbf{0}=0.\db{\alpha}$ and $\mathbf{\ominus\alpha_i}=-\alpha_i.\db{\alpha}$ so that $\mathbf {\alpha\oplus 0=\alpha}$ for all $\mathbf{\alpha}$ in $\db{\alpha}$ and
  $\mathbf{\alpha\,\,\oplus\,\,(\ominus\alpha)=0}$.
  From (\ref{e1}) and (\ref{e2}), we infer that $\left\{\alpha\right\}$ forms a one dimensional vector space. From experience, only similar quantities may be added, which leads to Axiom 1.
  \begin {axm}\label{a1}
  In scientific equations, the addition operation  between vectors belonging to different dimensional basis is not possible, so that if $\db{\alpha}\neq\db{\beta}$ then $\alpha_i.\db{\alpha}\,\oplus\,\beta_i.\db{\beta}$ is not defined.
  \end{axm}
  From Axiom \ref{a1}, it follows that $c_\alpha.\db{\alpha}$ is a 1-Dimensional vector space with the unique zero $0.\Mbl{\alpha_i}{i}$ which is not equatable with the zero vector of another space $c_{\beta}.\db{\beta}$ . In this sense, the dimensional bases $\db{\gamma_i}$ are orthogonal to each other.
\begin{defn}\label{d1}
The product operator $\otimes$ is a mapping $P$ such that\\
\begin{equation}\label{e3}
 P:\left\{c_\alpha.\db{\alpha},c_\beta.\db{\beta}\right\}\rightarrow c_\alpha .c_\beta.\left[\db{\alpha}\times\db{\beta}\right]
 \end{equation}
 and is defined as follows: 
$\left[c_{\alpha}.\db{\alpha}\right]^A \otimes \left[c_{\beta}.\db{\beta}\right]^B=(c_{\alpha})^A(c_{\beta})^B\Mb{(A.\alpha_i +B.\beta_i)}$\\
where $\left[\db{\alpha}\times\db{\beta}\right]=\Mb{(A.\alpha_i +B.\beta_i)}$, and $A,B \in F$ (but commonly restricted to $\Z$).The properties of $\otimes$ operator are as follows :
\begin{enumerate}
	\item if $B$ above is negative, then $c_{\beta} \neq 0.$
	\item it is symmetric, i.e. $\left[c_{\alpha}.\db{\alpha}\right]^A \otimes \left[c_{\beta}.\db{\beta}\right]^B=\left[c_{\beta}.\db{\beta}\right]^B\otimes \left[c_{\alpha}.\db{\alpha}\right]^A  $.
	\item it is associative, where for all elements $\Pi,\Theta,\Sigma$, the following obtains \\
	$(\Pi\otimes \Theta)\otimes \Sigma = \Pi\otimes (\Theta \otimes \Sigma)$.
	\item it is distributive over addition , i.e. \\
\begin{eqnarray}
	\left[c_{\alpha}.\db{\alpha}\right]^A \otimes \left[c_{\beta_1}.\db{\beta}\oplus c_{\beta_2}.\db{\beta} \right]&=&c_\alpha^{A} .c_{\beta_1}.\left[\db{\alpha}^A\times\db{\beta}\right]
	\oplus c_\alpha^{A} .c_{\beta_2}.\left[\db{\alpha}^A\times\db{\beta}\right]\nonumber \\
	&=&(c_\alpha^{A} .c_{\beta_1}+c_\alpha^{A}\label{e4} .c_{\beta_2}).\left[\db{\alpha}^A\times\db{\beta}\right].
	\end{eqnarray}
	\item the divisor operator $\oslash$ is defined such that $\oslash c_{\alpha}.\db{\alpha}=\otimes (c_{\alpha})^{-1}\Mbl{-\alpha_i}{i},\\
	c_{\alpha}\neq 0$. 
\end{enumerate}
%{\Fi}[3] {F_i^{\alpha,\,\beta,\ldots \, \gamma}\left(#1,\,,#2\,\ldots #3\,\right)}
\begin{defn}\label{d2}
A general mapping transformation $F_i^{\alpha,\,\beta,\ldots \,\gamma}$  which maps the domain $(c_\alpha.\db{\alpha}\, c_\beta.\db{\beta},\ldots \,c_\gamma.\db{ \gamma})$ to a range in $\db{R}$, $c_R.\db{R}$ in a series of operations involving only the $\oplus$ and $\otimes$ operators and scalar multiples is defined to be consistent. 
\end{defn}
\begin{axm}\label{a2}
Scientific equations are consistent mappings.
\end{axm}

From Definitions (\ref{d1}-\ref{d2}), consistent mappings are of the form 
\begin{equation}\label{e5}
	\Fi{c_\alpha.\db{\alpha}}{c_\beta.\db{\beta}}{c_\gamma.\db{ \gamma}}=\fisca{}{c_\alpha}{c_\beta}{c_\gamma}.\Mb{\fisca{\prime}{\alpha}{\beta}{\gamma}}
\end{equation}
where each $\sigma,\,(\sigma=\alpha,\beta,\ldots\gamma)$ are m-tuples $\sigma=(\sigma_1,\sigma_2,\,\ldots \sigma_m)$.
\begin{rem}
\begin{enumerate}
	\item[(a)]The mapping here involves the mapping of a collection of 1-D vector spaces $\left\{W_i\right\}\rightarrow V$ where the domain $V$ is also 1-D with basis $\db{v}$ where $\db{v}=\Mb{v_i}$.
\item[(b)]Transformation  (\ref{e5}) defines a scientific equation where $\fisca{}{c_\alpha}{c_\beta}{c_\gamma}\in F$  a member of the scalar field, representing a physical quantity with dimensions characterized by the functions $f_i^\prime$.
\end{enumerate}
\end{rem}
%{\fisca}[4] {f_i^{#1}(#2,\,#3,\,\ldots #4)}
\end{defn}
From (\ref{e3}) and (\ref{e4}), the field properties  under $(.,+)$ for a scalar function\\
 $\fisca{}{\alpha}{\beta}{\gamma}\rightarrow F$ is isomorphous to $(\otimes,\oplus)$ under the $1-1$ mapping $\alpha\rightarrow \alpha.\left[\mathrm{M^\alpha}\right]$ and so one can write 
\begin{equation}\label{e6}
\fisca{}{\alpha}{\beta}{\gamma}\left[\mathrm{M^{F'[\fisca{}{\alpha}{\beta}{\gamma}]}}\right]=
\fisca{}{\alpha.\left[\mathrm{M^\alpha}\right]}{\beta.\mathrm{M^\beta}}{\gamma.\mathrm{M^\gamma}}
\end{equation}
where the $(.,+)$ operators of the scalar function $f$  are replaced by the $(\otimes,\oplus)$  operators respectively in the R.H.S. of (\ref{e6}). The above is the reason why some people speak of  reduced units as being "`unitless"' \cite [p.199]{haile3}.
An example of a scientific equation for $f_i$ is the simple Lennard-Jones 12-6 potential,
\begin{equation} \label{e7}
 v^{LJ}(r,\epsilon,\sigma)=4\pi\epsilon \left(\left(\frac{\sigma}{r}\right)^{12}-\left(\frac{\sigma}{r}\right)^6\right)
 \end{equation}
 where the basis unit vectors are $1.\MLT{\alpha}{\beta}{\gamma}$ where $\mathrm{M,L\, and\, T}$ are the mass, length and time base unit symbols (e.g. kilogram , metre and seconds in S.I. units); $v^{LJ}$ is in this notation entirely unitless as all variables are members of the $\R$ field. From (\ref{e6}), the transformation function $F_{v^{LJ}}^{\alpha,\,\beta,\ldots \, \gamma}$  is in this case of the form
 \begin{eqnarray}
 \Fidwn {\epsilon.\MLT{0}{2}{-2}}  {r.\MLT{0}{1}{0}} {\sigma.\MLT{0}{1}{0}}{v^{LJ}}&& \nonumber\\
 =v^{LJ}(r,\epsilon,\sigma).\MLT{1}{2}{-2}&&\label{e8}
 \end{eqnarray}
 where the units of the potential are characterized by $\MLT{1}{2}{-2}$. Scientific equations are described relative to its dimensions; for two basis systems $\mathrm{M}$ and $\mathrm{M^\ast}$ describing the same physical phenomena, each having the same span length $m$ for its basis vectors , where for system $\mathrm{M}$, the basis is written $
{\rm M}_{\rm 1}^\alpha  {\rm M}_{\rm 2}^\beta   \ldots {\rm M}_{\rm m}^\gamma  
$ and similarly for the $\mathrm{M^\ast}$ basis, each dimensional symbol is related linearly, i.e.${\rm M}_{\rm i}^{\rm *} {\rm  = \lambda }_{\rm i} {\rm M}_{\rm i} $. This leads to Axiom 3.
  
\begin{axm}\label{a3}
Scientific equations describing phenomena relative to the basis\\ $1.\Mb{}$ may be described relative to another basis $1.\Mb{\ast}$ through fixed scaling parameters $\lambda_i$ where 
\begin{equation}\label{e9}
\left(\prod_{i=1}^m
(\lambda _i^{\alpha _i })\right) .\Mb{\alpha_i}=1.\Mb{\ast \,\,\alpha_i}.
\end{equation}
\end{axm}
The vector spaces are therefore linearly dependent. Scientifically, the choice of units cannot led to different physical phenomena through the scientific equations describing the system trajectory, which is expressed in Axiom 4.
\begin{axm}\label{a4}
The mappings describing scientific laws  in scientific equations are independent of chosen unit dimensions, so that for any scientific equation 
\begin{equation}\label{e10}
\Fi{c_\alpha.\db{\alpha}}{c_\beta.\db{\beta}}{c_\gamma.\db{ \gamma}}
=\Fi{c^{\ast}_\alpha.\db{\alpha}^\ast}{c^{\ast}_\beta.\db{\beta}^\ast}{c^{\ast}_\gamma.\db{ \gamma}^\ast}
\end{equation} 
\end{axm}
 
Axiom \ref{a4} and (\ref{e5}) yields for the R.H.S. of (\ref{e10}) the following 
 \begin{eqnarray}
 \Fi{c^{\ast}_\alpha.\db{\alpha}^\ast}{c^{\ast}_\beta.\db{\beta}^\ast}{c^{\ast}_\gamma.\db{ \gamma}^\ast}
 &=&\fisca{}{c_\alpha^\ast}{c^{\ast}_\beta}{c^{\ast}_\gamma}.\Mb{\ast \,\,\,\nonumber \fisca{\prime}{\alpha_i}{\beta_i}{\gamma_i}}\\ 
&=&f_i^\ast.\Mb{\ast\,\,\, f_i^{\prime}}. \label{e11}
 \end{eqnarray}
 Here we write $f_i^\ast=f_i$. Axiom \ref{a3} and (\ref{e11}) implies 
 \begin{eqnarray}
 \Fi{c^{\ast}_\alpha\lambda_\alpha.\db{\alpha}}{c^{\ast}_\beta \nonumber \lambda_\beta.\db{\beta}}{c^{\ast}_\gamma.\lambda_\gamma\db{ \gamma}} && \\  
 =\fisca{}{c_\alpha^\ast}{c^{\ast}_\beta}{c^{\ast}_\gamma}.\left(\prod \lambda_i^{{f^\prime}(\alpha_i,\,\beta_i,\ldots \, \gamma_i)}\right).\Mb{{f^\prime}(\alpha_i,\,\beta_i,\ldots \, \gamma_i)}.&& \label{e12}
 \end{eqnarray}
 But,
 \begin{equation}\label{e13}
 \Fi{c^{}_\alpha.\db{\alpha}}{c^{}_\beta.\db{\beta}}{c^{}_\gamma.\db{ \gamma}}   
 =\fisca{}{c_\alpha}{c_\beta}{c_\gamma}.\Mb{{f^\prime}(\alpha_i,\,\beta_i,\ldots \, \gamma_i)}.
 \end{equation}
 Thus we have 
\begin{equation}\label{e14}
\fisca{}{c_\alpha^\ast}{c^{\ast}_\beta}{c^{\ast}_\gamma}.\prod
\lambda_i^{{f}(\alpha_i,\,\beta_i,\ldots \, \gamma_i)}={f}(\alpha_i,\,\beta_i,\ldots \, \gamma_i).
\end{equation}
Axiom \ref{a3} and (\ref{e9}) give 
\begin{eqnarray}\label{e15}	c_\alpha.\db{\alpha}&=\frac{c_\alpha}{\Piprod{}\left(\lambda_i^{\alpha_i}\right)}\Mbl{\ast}{\alpha,\,i}&=\frac{c_\alpha}{\lambda_\alpha}\Mbl{\ast}{\alpha,\,i}=c_{\alpha}^\ast.\Mbl{\ast}{\alpha}
\end{eqnarray}
or,
\begin{eqnarray}\label{e16}
	c_{\alpha}^\ast&=\frac{c_\alpha}{\Piprod{}\left(\lambda_i^{\alpha_i}\right)}&=\frac{c_\alpha}{\lambda_\alpha}.
\end{eqnarray}
From the above, the star operator ($\ast$)corresponding to a change of unit basis $1.\M{}$ to $1.\M{^\ast}$ and any scalar function with variables $c_\alpha$ linked to the vector $c_\alpha.\M{_\alpha}$ may be written
\begin{eqnarray}\label{e17}
	{f^\ast}(c_\alpha,\,c_\beta,\ldots \, c_\gamma)&=&{f}(c_\alpha^\ast,\,c_\beta^\ast,\ldots \, c_\gamma^\ast)\nonumber \\ 
	&=&{f}(\frac{c_\alpha}{\lambda_\alpha},\,\frac{c_\beta}{\lambda_\beta},\ldots \,\frac {c_\gamma}{\lambda_\gamma})
\end{eqnarray}
with $(\lambda_\alpha,\lambda_\beta\ldots\lambda_\gamma)$ given in (\ref{e16}).
\section{Discussion and verifications}
\subsection{(a)some standard applications}
In the laboratory $\M{}$ basis, the interparticle  potential has sometimes   been modeled according to (\ref{e7}); a change in the unit basis implies converting the laboratory frame of units  given in (\ref{e7}) to another $\M{^\ast}$ with  form  given by (\ref{e11}). In this situation,let there be a unit basis of length such that 
$\sigma.\M{_L}=1.\M{^{\ast}_L}$ or in detail $\sigma\M{_L}=\sigma.\MLT{0}{1}{0}=1.\M{_L^{\ast}}=1.\MLT{\ast 0}{\ast 1}{\ast 0}$. Further, let the energy scale as $\epsilon.\M{_E}=1.\M{^\ast_E}$. In (\ref{e7}), $\sigma$ and $r$ are linked to $\M{_L}$; the transforming operator on $v^{LJ\ast}=v^{LJ}\left(\sigma^\ast,\epsilon^\ast,r^\ast\right)$ yields
\begin{eqnarray}
	v^{LJ\ast}=4\pi\epsilon^\ast \left(\left(\frac{\sigma^\ast}{r^\ast}\right)^{12}-\left(\frac{\sigma^\ast}{r^\ast}\right)^6\right)\nonumber \\
	=4\pi \left(\left(\frac{1}{r^\ast}\right)^{12}-\left(\frac{1}{r^\ast}\right)^6\right)
\end{eqnarray}
which is a reduced potential used in simulations. In the above case,  the bases for energy and mass  are $\M{_E}=\MLT{1}{2}{-2}$ and $\M{_M}=\MLT{1}{0}{0}$ respectively. If 3 scales are chosen for the  $\lambda$'s, e.g. $(\epsilon,\sigma,m)$, then other quantities are fixed relative to it. From  the kinetic energy, we allow $\epsilon=\frac{1}{2}mv^2$, then  the scaling for the velocity is also determined, and so $v^2$ is  a fix quantity. We require $v^\prime$, the $\lambda$ parameter for velocity scaling. Applying (\ref{e17}), the results are $\epsilon^\ast=1=\frac{1}{2} \left(\frac{v}{v^\prime}\right)^2$ or $v^{\prime\,2}=\frac{v^2}{2}=\frac{\epsilon}{m}$ which results in the scaled velocity $v^\ast=v\sqrt{\frac{m}{\epsilon}}$ in accordance with standard results \cite{haf1}.
\subsection{(b)the temperature parameter}
The above theory represents scientific numbers in terms of $c_\alpha.\MLT{\alpha_1}{\alpha_2}{\alpha_3}$, and clearly, if a unit cannot be expressed in terms of the defined basis given here, then absolutely \textit{no} scaling parameter $\lambda$ exists for that quantity, and in particular $(a.c)^\ast\neq a^\ast . c^\ast$ since the isomorphic properties of the above-mentioned operators only apply to  quantities with an associated dimension $c_\alpha .\M{_\alpha}$; for example scaling with fixed parameters $(m,\sigma,\epsilon)$ means that the  velocity scaling parameter $\lambda_v$ is defined such that $v^\ast=\frac{v}{\lambda_v}$ is $\left(\frac{\epsilon}{m}\right)^{1/2}$, where the dimension of this quantity, as well as the scaling parameters all have the form $c_\alpha.\M{_\alpha}$. Temperature (and other associated properties) do not emerge directly from mechanical analysis - where the basis dimensions are defined - but from theoretical models or definitions, and in particular cannot be defined in the form $c_\alpha .\M{_\alpha}$ . It is therefore misleading to suppose  that a $\lambda_T$ divisor must exist for this quantity such that $T^\ast=\frac{T}{\lambda_T}$ which is  often  suggested e.g.  \cite [Table 1.]{haf1}; the actual implication, as shown below is the definition of another temperature scale $\mathcal{T}=k_B T$ where $\mathcal{T}^\ast=\left(\frac{k_BT}{\epsilon}\right)$. Likewise the Boltzmann coupling parameter $k_B$ has a unit which is reciprocal to temperature, implying that it cannot be expressed in the form $c_\beta .\M{_\beta}$ and hence this parameter cannot be reduced in isolation by setting "`$k_B=1$, so that the MD unit of temperature is also defined"' \cite[pp.15-16]{rap2}. This assumption of the fundamental mechanical autonomy of the coupling parameter and temperature is however rather standard and pervasive \cite[p.200]{haile3}and  in nearly all cases, the standard  assignments are correct, which is explained below.

In MD simulations, the temperature is determined via the classical equipartition theorem (which is known from experience and quantum mechanics not to obtain at lower temperatures for the free vibrational and rotational modes) for all temperatures from the mean kinetic energy of translation where 
\begin{equation}   \label{e19}
\left(k_BT\right)	=\frac{1}{ND}\left\langle \sum_{i=1}^{N-D} mv^2_i \right\rangle
\end{equation}
and where $D=2,3$ for two and three dimensional systems respectively for an $N$ particle system where the angle brackets denotes some chosen averaging algorithm that is thought to approximate the outcome if the $P$ density function mentioned below is used in an exact evaluation. Investigators claim that due to "`conservation of momentum"', there are $D$ degrees of freedom that must not be counted in (\ref{e19}),\cite[p.16 and pp.46-47 respectively]{rap2,til4}.
The probability density function $P$ over the $(\mathbf{p,q})$ momentum and positional coordinates  for the above averaging process is $P=\exp{\frac{-\mathcal{H}(\mathbf{p,q})}{k_BT}/Q}$ with $Q=\int_{\partial V}\exp{\frac{-\mathcal{H}(\mathbf{p,q})}{k_BT}}\mathbf{dp\:dq}$ defined as the phase integral, which is the analog of the quantum partition function for the canonical distribution.$\mathcal{H}$ is the Hamiltonian , written in most classical simulations with the momentum and potential coordinates separated viz. 
$\mathcal{H}=\sum_{i=1}^N\frac{p_i^2}{2m_i}+V(\mathbf{q})$.
 Using this $P$ density, each particle yields $<p^2_i/2m>=\frac{D}{2}(k_BT)$ , where clearly the definition of $T$ is the consequence of stochastic averaging; it is not a fundamental mechanical quantity associated with a basis dimension. Extending this result to  $N$ particles  is thought to yield (\ref{e19}) , which is routinely used to compute the temperature in simulations. Averaging the kinetic energy  of $\mathcal{H}$ using $P$   (which already has the constraint over the entire ensemble of total energy conservation - from where the $\beta$ parameter from the Lagrange multiplier represents the $k_BT$  term and the $\alpha$ term from the Lagrange multiplier refers to total particle conservation which is featured in the phase integral or partition function)  does not indicate any such reduction of degrees of freedom. Indeed, from the Gibbs' postulate of the equivalence of ensemble average  to that of the time average of a particular system, one can view each particle as a system, from which we can expect that the time average of the mean kinetic energy  of any given particle would equal $\frac{D}{2} k_BT$ with the same $k_BT$ as given in (\ref{e19}) for the entire system of which it is a part. Further, the supposed reduction of the number of degrees of freedom implies that no temperature can exist for a  system comprising of $D$ or less particles, and this is patently absurd, for it is eminently feasible to conceive and implement thermostats in MD for such systems. Hence the  proper form of the temperature must be  derived from the probability density function and/or Gibbs' postulate with the average energy per particle given as $\frac{D}{2}k_BT$ to yield 
\begin{equation}   \label{e20}
\left(k_BT\right)	=\frac{1}{ND}\left\langle \sum_{i=1}^{N} mv^2_i \right\rangle
\end{equation}
if the classical kinetic energy is used as an indicator. The error reported in studies would be due to the degree of difference between (\ref{e19}) and (\ref{e20}).

From  the vector space calculus here, $\left( k_BT \right)$ can be expressed as $c_\beta .\M{_\beta}$ , but $\left( k_BT \right)^\ast\neq k_B^\ast\,T^\ast$ since the temperature  and coupling constant is not separately definable.Hence another temperature parameter $\mathcal{T}$ may be defined where $\T=k_BT$, so that $\left(\T\right)^\ast =\left(k_BT\right)^\ast$ and hence $\T^\ast=(k_BT)^\ast	=\left(\frac{1}{ND}\left\langle \sum_{i=1}^{N} m^\ast v^{\ast 2}_i \right\rangle\right)$ and in particular $\T^\ast=\left(\frac{k_BT}{\epsilon}\right)$, which is the standard assignment \cite {haf1}, which leads to standard  and consistent results, provided it is understood that $\T=k_BT$ and \textit{not}  $\T=T$, which is the normal understanding, where it is assumed \cite {rap2} that $k_B^\ast=1$;  even with this unfortunate assumption, together with the autonomy of variables, the correct results are derived because of the following equation sequence: 
$k_B^\ast T^\ast=\left(k_B T\right)^\ast =\frac{k_BT}{\epsilon}$ or $T^\ast=\frac{k_BT}{\epsilon k_B^\ast}=\frac{k_BT}{\epsilon }$ which is the fortuitously  correct result with $k_B^\ast=1$ but with the incorrect algebraic assumptions, since $k_B$ is not independent, it cannot be arbitrarily set to a value.

There are clear-cut consequences that follow from whether (a) an independently scaled $k_B^\ast$ exists, or, (b) where this is not the case. In the simulation of entropic quantities based on  the Boltzmann postulate for entropy $S$ given by  $S=k_B \ln W$ , case (a) implies $S^\ast=\ln W$, or a direct determination is possible for the entropy; (b) suggests a work-around, such as scaling equations as (i)$S^\prime = \left(S/{k_B}\right)$ or (ii) $E^\prime =\left(TS\right)$ in constant temperature studies. The techniques used in entropic  studies use(i) and  (ii) or a variant of (ii) through the  determination first of the chemical potential scaled in $k_BT$ units \cite [pp.246-249]{haile3}, or in coupling methods (utilizing thermodynamical integration), $S/k_B$ is the variable that is scaled in simulations (method (i) ) \cite [p.260]{haile3}.It is proposed here that simulations utilize variants of either (i) or (ii)  above as a direct consequence of (b), that an independent $k_B$ and $T$ (or $k_B^\ast$ and $T^\ast$) does not exist for  scaling.

\textbf{Acknowledgment}: I thank Thomas Bier (Institute of Mathematical Sciences, U.M.) for discussions.

%%%%%%%%%%%%%%%%%%%%%%%%%%%%%%%%%%%%%%%%%%%%%%%%%%%%%%%%%%%%%%%%%%%%%%%%%%%%%%%%%%%%

{}
\end{document}